\documentclass[9pt,final,journal]{IEEEtran}
\usepackage[latin1]{inputenc}
\usepackage{amsmath,amssymb}
\usepackage{graphicx}
\usepackage[tight]{subfigure}
\usepackage{enumerate}
\newtheorem{defin}{Definition}
\newtheorem{teo}{Theorem}
\newtheorem{lema}{Lemma}
\newtheorem{remark}{Remark}

\newtheorem{prop}{Proposition}

\newtheorem{as}{Assumption}

\def\K{\mathcal{K}}
\def\Ki{\K_{\infty}}
\def\KL{\mathcal{KL}}
\def\R{\mathbb{R}}
\def\N{\mathbb{N}}
\def\S{\mathcal{S}}
\def\T{\mathcal{T}}
\def\U{\mathcal{U}}
\def\F{\mathcal{F}}
\def\dom{{\rm dom}}
\def\mer{\hfill $\circ$}
\def\sat{\mathrm{sat}}
\def\sign{\mathrm{sign}}
\def\diag{\mathrm{diag}}
\def\comp{{\scriptstyle\,\circ}\,}

\title{A Characterization of Integral ISS\\ for Switched and Time-varying Systems}
\author{H. Haimovich and J.L. Mancilla-Aguilar  
 \thanks{H. Haimovich is with CIFASIS, CONICET-UNR, Ocampo y Esmeralda, 2000 Rosario, Argentina. {\texttt{haimovich@cifasis-conicet.gov.ar}}}%
\thanks{J.L. Mancilla-Aguilar is with Departamento de Matem\'atica, Instituto Tecnol\'ogico de Buenos Aires, Avda. Madero 399, Buenos Aires, Argentina. \texttt{jmancill@itba.edu.ar}}
\thanks{Work partially supported by ANPCyT grants PICT 2013-0852 and PICT 2014-2599, Argentina.}%
}

\begin{document}
\maketitle

\begin{abstract}
  Most of the existing characterizations of the integral input-to-state stability (iISS) property are not valid for time-varying or switched systems in cases where converse Lyapunov theorems for stability are not available. This note provides a characterization that is valid for switched and time-varying systems, and shows that natural extensions of some of the existing characterizations result in only sufficient but not necessary conditions. The results provided also pinpoint suitable iISS gains and relate these to supply functions and bounds on the function defining the system dynamics.
\end{abstract}


\section{Introduction}
\label{sec:introduction}

Input-to-state stability (ISS) \cite{sontag_tac89} and integral-ISS (iISS) \cite{sontag_scl98} are arguably the most important and useful state-space based nonlinear notions of stability for systems with inputs. The ISS property gives a state bound that is the sum of a decaying-to-zero term whose amplitude depends only on the initial state, and a term depending (nonlinearly) only on the input bound. The difference in the iISS property lies on the input-dependent term, which is a (nonlinear) function of an input energy bound, instead of an input bound.

For time-invariant systems, several characterizations of both the ISS and iISS properties exist (see \cite{sonwan_scl95,sonwan_tac96,libshi_tac15} for ISS and \cite{sontag_scl98,angson_tac00,angson_dc00} for iISS). Among the different characterizations of these properties, perhaps the most practical ones are those based on ISS- \cite{sonwan_scl95} or iISS- \cite{angson_tac00} Lyapunov functions. Indeed, since each of these properties is known to be equivalent to the existence of the respective type of Lyapunov function, there is no loss of generality in focusing on the obtention of such functions. 
Results that ensure that an ISS or iISS system admits the corresponding type of Lyapunov function heavily rely on converse Lyapunov theorems for stability \cite{linson_jco96}, since both ISS and iISS imply global asymptotic stability. 

As for time-varying systems, some Lyapunov characterizations of ISS exist in both uniform \cite{edwlin_cdc00} and non-uniform flavors \cite{kartsi_tac04,linwan_ifacwc05}. All of these works assume that the function $f$ defining the system dynamics, $\dot x = f(t,x,u)$, is (at least) continuous. To the best of our knowledge, no useful characterizations of the iISS property nor iISS-Lyapunov converse theorems exist in this case.

This paper deals with time-varying systems, especially with switched systems \cite{liberzon_book03}, and focuses on iISS that is uniform over some given set of switching signals ---the set of \emph{admissible} switching signals--- (see Section~\ref{sec:prel} for the precise definition). In this setting, the situation can be fairly different depending on the properties of this set. For example, when the set of admissible switching signals coincides with the set of all switching signals, i.e. under arbitrary switching, and the subsystems are time-invariant then the Lyapunov characterizations available for ISS and iISS carry over to the switched system with little change \cite{mangar_scl00,mangar_scl01}. However, to the best of the authors' knowledge, all of the existing converse theorems employed to derive Lyapunov characterizations of the ISS or iISS properties break down if the set of admissible switching signals is not closed under concatenations (i.e. when piecing together two admissible switching signals does not necessarily result in another admissible switching signal). As a consequence, when the set of admissible switching signals is not closed under concatenations, such Lyapunov characterizations cannot be derived by following known techniques and, as we will illustrate along this paper, it is likely that no such Lyapunov characterization is possible. Sets of switching signals not closed under concatenations have not only theoretical but also practical interest (e.g. in the analysis of stability of switching converters \cite{denhai_auto16} and supervisory control \cite{liberzon_book03}).

In this context, the main contribution of the current paper is to provide a characterization of iISS for switched systems with any set of admissible switching signals. More specifically, we will show that a switched system is iISS uniformly with respect to a given set of switching signals if and only if the system satisfies a uniformly bounded energy bounded state \cite{angson_dc00} property and a 0-input global asymptotic stability property, both uniformly with respect to the given set of switching signals. This characterization of iISS was originally developed in \cite{angson_dc00} for time-invariant systems. The corresponding proof in \cite{angson_dc00} is based on a converse Lyapunov argument which is not valid in the setting considered here. Hence, the proof in the current paper is, to the best of our knowledge, completely novel, even for the case of non-switched time-varying systems. A second contribution is to show that characterizations of iISS based on dissipation inequalities and appropriate detectability conditions \cite{angson_tac00} (see also \cite{wanwei_tac08,jayrya_tac10}) become only sufficient but not necessary in the setting considered. In the case of time-varying and switched systems, a natural extension of the weak zero-detectability property in \cite{angson_tac00} is given by the output-persistent excitation property \cite{leejia_tac08,leetan_tac15} (see Section~\ref{sec:dissipativity-iiss}). A third contribution is that our results also pinpoint iISS gains, and relate these to supply functions and bounds on the function defining the system dynamics, thus extending also some results of \cite{jayrya_tac10} regarding the bounded-energy-input/convergent-state property. 

The results in the current paper are novel even for time-varying non-switched systems, and apply to switched systems of most general forms, having time-varying (not necessarily continuous with respect to time) and nonlinear subsystems, and imposing mild conditions on the function defining the system dynamics. In addition, our results can be interpreted as conditions for iISS of a given arbitrary family of time-varying systems, where the iISS estimate is uniform over all the systems in the family. In this regard, our results are not limited to the case where the family of time-varying systems arises from the consideration of a switched system (see Remark~\ref{rem:families} in Section~\ref{sec:unifiiss}). However, we chose to keep the formulation in terms of switched systems due to its natural application to the latter type of systems.

%
%


The remainder of this note proceeds as follows. This section ends with a brief description of the notation employed. Section~\ref{sec:prel} describes the system considered, states the standing assumptions, and precisely defines the properties employed. Our main results are given in Sections~\ref{sec:char-iiss} and~\ref{sec:dissipativity-iiss}. Section~\ref{sec:char-iiss} provides a characterization of iISS and Section~\ref{sec:dissipativity-iiss} proves that conditions based on dissipativity are only sufficient but not necessary. Examples are provided in Section~\ref{sec:examples} and conclusions drawn in Section~\ref{sec:concl}.

\textbf{Notation.} $\N$, $\R$, $\R_{>0}$ and $\R_{\ge 0}$ denote the natural numbers, reals, positive reals and nonnegative reals, respectively. $|x|$ denotes the Euclidean norm of $x \in \R^p$. Vector or matrix transposition is denoted by~$'$. For any $m\in \N$, $\U_m$ denotes the set of all the Lebesgue measurable and locally essentially bounded functions $u:\R_{\ge 0}\to \R^m$. We write $\alpha\in\K$ if $\alpha:\R_{\ge 0} \to \R_{\ge 0}$ is continuous, strictly 
increasing and $\alpha(0)=0$, and $\alpha\in\Ki$ if, in addition, $\alpha$ is unbounded. We write $\beta\in\KL$ if $\beta:\R_{\ge 0}\times \R_{\ge 0}\to \R_{\ge 0}$, $\beta(\cdot,t)\in\Ki$ for any $t\ge 0$ and, for any fixed $r\ge 0$, $\beta(r,t)$ monotonically decreases to zero as $t\to \infty$. By `index set', we mean an arbitrary nonempty set, not necessarily finite nor countable.

\section{Preliminaries}
\label{sec:prel}

\subsection{Time-varying and switched systems with inputs}
\label{sec:tvswsys}

Consider a time-varying switched system with inputs $u$, of the form
\begin{align}
  \label{eq:1}
  \dot x = f(t,x,u,\sigma)
\end{align}
where $t\in\R_{\ge 0}$, $x(t)\in\R^n$, $u \in \U_m$ and $\sigma:\R_{\ge 0}\to \Gamma$, with $\Gamma$ an index set, is a switching signal, i.e. it is piecewise constant (having a finite number of discontinuities in every bounded interval) and right continuous. We assume that $f:\R_{\ge 0}\times \R^n\times \R^m\times \Gamma\to \R^n$
satisfies $f(t,0,0,i)=0$ for all $t\ge 0$ and all $i\in\Gamma$, and that $f(t,\xi,\mu,i)$ is Lebesgue measurable in $t$ for fixed $(\xi,\mu,i)$ and continuous in $(\xi,\mu)$ for fixed $t$ and $i$. The following is our main technical assumption.
\begin{as} \label{ass:A}
 $f$ in \eqref{eq:1} satisfies
 \begin{enumerate}[C1)]
  \item \label{item:bound} There exist $\gamma \in \K$ and a nondecreasing function $N:\R_{\ge 0}\to \R_{>0}$ such that 
 $|f(t,\xi,\mu,i)|\le 
N(|\xi|)(1+\gamma(|\mu|))$ for all $t\ge 0$, all $\xi \in \R^n$, all $\mu\in \R^m$ and all $i\in \Gamma$.
\item For every $r>0$ and $\varepsilon >0$ there exists $\delta>0$ such that for all $t\ge 0$ and $i\in \Gamma$, 
$|f(t,\xi,\mu,i)-f(t,\xi,0,i)|<\varepsilon$ if $|\xi|\le r$ and $|\mu|\le\delta$.\label{item:fcont}
 \item  \label{item:lips} $f(t,\xi,0,i)$ is locally Lipschitz in $\xi$, 
 uniformly in $t$ and $i$.\mer
\end{enumerate} 
\end{as}
\begin{remark}
  Existing characterizations of the integral ISS property for systems of the form $\dot x = f(x,u)$ (cf. \cite{sontag_scl98,angson_tac00,angson_dc00}) usually assume that $f(x,u)$ is locally Lipschitz. We emphasize that we do not require that $f$ in (\ref{eq:1}) satisfy any additional Lipschitzity requirement other than that in C\ref{item:lips}) of  Asumption \ref{ass:A}. As a consequence, solutions to (\ref{eq:1}) are not necessarily unique. \mer
\end{remark}
Assumption~\ref{ass:A} is indeed guaranteed to hold, for example, when $f$ in (\ref{eq:1}) satisfies a local uniform Lipschitz condition (see Lemma~\ref{lem:loc-lips} below), and also for control-affine systems where $f(t,\xi,\mu,i) = f_0(t,\xi,i)+g(t,\xi,i) \mu$ with $f_0(t,\xi,i)$ Lebesgue measurable in $t$, locally Lipschitz in $\xi$ uniformly in $t$ and $i$, and $f_0(t,0,i)\equiv 0$, and $g(t,\xi,i)$ is Lebesgue measurable in $t$, continuous in $\xi$, and for every $r>0$ there exists $N\ge 0$ such that for all $t\ge 0$ and $i\in \Gamma$, $|g(t,\xi,i)|\le N$ if $|\xi|\le r$. The proof of the following result is given in the Appendix.
\begin{lema} \label{lem:loc-lips} If $f(t,\xi,\mu,i)$ is locally Lipschitz 
in $(\xi,\mu)$ uniformly in $t$ and $i$, and $f(t,0,0,i)=0$ for all $t\ge 0$ and all $i\in\Gamma$, then Assumption \ref{ass:A} holds.
\end{lema}

We will employ $\S$ to denote a set of (admissible) switching signals. Given $t_0\ge 0$, $u\in \U_m$ and $\sigma\in \S$, we denote by $\T(t_0,u,\sigma)$ the set of maximally defined solutions $x$ of (\ref{eq:1}) corresponding to $u$ and $\sigma$ such that $t_0\in \dom \:x$, where $\dom\:x$ denotes the interval of definition of $x$. We say that $(\ref{eq:1})$ is forward complete with respect to (w.r.t.) $\S$ if for all $t_0\ge 0$, $u\in \U_m$, $\sigma\in \S$ and $x\in \T(t_0,u,\sigma)$, $[t_0,\infty)\subset \dom\:x$.

\subsection{Uniform integral ISS}
\label{sec:unifiiss}

For the switched system (\ref{eq:1}) and the set $\S$ of switching signals, we will consider an integral input-to-state stability property that is uniform over switching signals in $\S$. This property is thus an extension of the one introduced in \cite{sontag_scl98}.
\begin{defin}
  \label{def:iiss}
  System (\ref{eq:1}) is said to be iISS w.r.t. $\S$ if it is forward complete w.r.t. $\S$ and there exist $\beta\in\KL$, and $\rho$ and $\chi \in\K$ (the latter will be referred to as an iISS gain) such that the estimate (\ref{eq:xestiiss}) holds for all $t\ge t_0\ge 0$, all $u\in \U_m$, all $\sigma\in \S$ and all $x\in \T(t_0,u,\sigma)$.
  \begin{align}
    \label{eq:xestiiss}
    |x(t)| \le \beta(|x(t_0)|,t-t_0) + \rho\left(\int_{t_0}^t \chi(|u(\tau)|) d\tau \right).
  \end{align}
\end{defin}
\begin{remark}
  \label{rem:families}
  We may equivalently formulate the problem considered as follows. For each $\sigma\in\S$ define $f_\sigma : \R_{\ge 0} \times \R^n \times \R^m \to \R^n$ via $f_\sigma(t,\xi,\mu) := f(t,\xi,\mu,\sigma(t))$, with $f$ in (\ref{eq:1}). Consider the family $\F$ of time-varying systems given by $\F:= \{f_\sigma : \sigma\in\S\}$. Then, we aim at characterizing iISS so that the estimate (\ref{eq:xestiiss}) holds uniformly for every system in the family $\F$. Since we do not impose any additional restrictions on the set $\S$, we may also consider arbitrary families $\F$ of time-varying systems, not only those arising from a switched system, i.e. we may consider $\S$ to be an arbitrary index set and rewrite the above assumptions in terms of $f_\sigma$ instead of $f$, so that the assumptions on $f_\sigma$ hold uniformly over every possible $\sigma\in\S$. For example, instead of assuming C1) 
  we would require C1') there exist $\gamma \in \K$ and $N$ nondecreasing so that $|f_\sigma(t,\xi,\mu)|\le N(|\xi|)(1+\gamma(|\mu|))$ for all $t\ge 0$, $\xi\in \R^n$, $\mu \in \R^m$ and $\sigma\in\S$. We emphasize that C1') is even weaker than C1) in the case of switched systems, but keep our assumptions as above for the sake of simplicity. 
  \mer
\end{remark}
Besides the uselfulness of the iISS property for describing the qualitative behaviour of the solutions of (\ref{eq:1}), the computation of an iISS gain is pertinent to the stability analysis of interconnected systems which contain iISS subsystems, and to the robustness analysis of closed-loop systems (see \cite{arcang_siamjco02}, \cite{itojia_cdc06}).
Given $\chi \in \K$ and $u\in \U_m$, let $\|u\|_{\chi}:=\int_{0}^{\infty} \chi(|u(\tau)|) d\tau$ and $\U_m^\chi:=\{u\in \U_m:\;\|u\|_{\chi}<\infty\}$. However, note that $\|\cdot\|_{\chi}$ is not necessarily a norm in $\U_m^\chi$.

\begin{remark} 
\label{rem:iiss-norm}
  Due to causality and the Markov property, an equivalent definition of iISS w.r.t. $\S$ is obtained if $\int_{t_0}^t \chi(|u(\tau)|) d\tau$ in 
(\ref{eq:xestiiss}) is replaced by $\|u\|_{\chi}$.\mer
\end{remark}

An equivalent definition of iISS is provided by Lemma~\ref{rem:iiss-charact} below. This lemma can be proved by techniques analogous to those in the proof of Lemma~2.7 of \cite{sonwan_scl95}, and considering Remark \ref{rem:iiss-norm}.
\begin{lema}
  \label{rem:iiss-charact}
 System~(\ref{eq:1}) is iISS w.r.t. $\S$ with iISS gain $\chi$ if and only if it is forward complete w.r.t. $\S$ and the following conditions hold with $\|u\|=\|u\|_{\chi}$:
    \begin{enumerate}[i)]
    \item For every $T>0$, $r>0$ and $s>0$ there exists $C>0$ such that every  
    $x\in \T(t_0,u,\sigma)$, with $t_0\ge 0$, $u\in \U_m$ and $\sigma\in \S$ such that $\|u\|\le s$ and $|x(t_0)|\le r$,  
    satisfies $|x(t)|\le C$ for all $t\in [t_0, t_0+T]$. \label{item:bound-iss}
\item For each $\epsilon>0$ there exists $\delta>0$ such that every $x\in \T(t_0,u,\sigma)$, with $t_0\ge 0$, $u\in \U_m$ and $\sigma\in \S$ such that 
$\|u\|\le \delta$ and $|x(t_0)|\le \delta$, 
    satisfies $|x(t)|\le\epsilon$ for all $t\ge t_0$. \label{item:gs-iss}
\item There exists $\nu\in\K$ such that, for any
      $r\ge \epsilon>0$, there is a $T>0$ so that for every $x\in \T(t_0,u,\sigma)$, with $t_0\ge 0$, $u\in \U_m$ and $\sigma\in \S$ such that 
$|x(t_0)|\le r$, then\label{item:ua-iss}
      \begin{align*}
        |x(t)| \le \epsilon + \nu(\|u\|)\quad\forall t\ge t_0 + T.
      \end{align*}
    \end{enumerate}
\end{lema}
\subsection{Bounded-energy-input/convergent-state property}

We will also consider the following convergence property, providing a natural extension of the one considered in \cite{jayrya_tac10} for time-invariant systems. 
\begin{defin}
We say that (\ref{eq:1}) has the bounded-energy-input convergent-state property (BEICS) w.r.t. $\|\cdot\|_{\chi}$ and $\S$ ($\chi$-BEICS w.r.t. $\S$, for short) if for every $x\in \T(t_0,u,\sigma)$ with $t_0\ge 0$, $u\in \U_m^{\chi}$ and $\sigma\in \S$, $x(t)\to 0$ as $t\to \infty$. 
\end{defin}
The BEICS property is useful for establishing asymptotic stability of cascade systems \cite{arcang_siamjco02}. The following fact can be proved in the same way as Proposition 6 in \cite{sontag_scl98}.
\begin{prop} 
  \label{prop:iISSimpliesBEICS} 
  Suppose that (\ref{eq:1}) is iISS w.r.t. $\S$ with iISS gain $\chi$. Then (\ref{eq:1}) is $\chi$-BEICS w.r.t. $\S$.
 \end{prop}

\subsection{The zero-input system}
\label{sec:zeroinputsys}

In the sequel, we will refer to the 0-input system corresponding to (\ref{eq:1}). We will employ $\mathbf{0}$ to denote the input $u\in\U_m$ such that $u(t) = 0$ for all $t\ge 0$. The 0-input system is thus the system defined by $\dot x = f(t,x,0,\sigma)$, and according to the definitions above, $\T(t_0,\mathbf{0},\sigma)$ is the set of its maximally defined solutions corresponding to the switching signal $\sigma$ such that $t_0 \in \dom\: x$. The following stability property of the 0-input system will be required in the sequel.
\begin{defin}
  \label{def:0guas}
  System~(\ref{eq:1}) is said to be zero-input globally uniformly asymptotically stable (0-GUAS) w.r.t. $\S$ if there exists $\beta\in\KL$ such that every $x\in\T(t_0,\mathbf{0},\sigma)$, with $t_0\ge 0$ and $\sigma\in \S$, verifies
  \begin{align}
    \label{eq:xest0guas}
    |x(t)| \le \beta(|x(t_0)|,t-t_0)\quad\forall t\ge t_0 \ge 0.
  \end{align}
\end{defin}
From Definitions~\ref{def:iiss} and~\ref{def:0guas}, it is clear that iISS w.r.t. $\S$ implies 0-GUAS w.r.t. $\S$.

\section{Characterization of iISS} 
\label{sec:char-iiss}

In this section we provide a characterization of the iISS w.r.t. $\S$ property. This characterization essentially is an extension of the equivalence 1 $\Longleftrightarrow$ 2 in Theorem~1 of \cite{angson_dc00}. Our approach is substantially different, however, because converse Lyapunov theorems cannot be successfully applied in the setting considered. Our results also have the advantage of pinpointing suitable iISS gains. We thus introduce the following definition, which is a natural extension of the corresponding one in \cite{angson_dc00}.
\begin{defin} 
  \label{def:ubebs}
  System (\ref{eq:1}) is uniformly bounded energy bounded state (UBEBS) w.r.t $\S$, if for some functions $\alpha_1, \alpha_2, \alpha \in \K$ (the latter will be referred to as an UBEBS gain), and some $c\ge 0$, the estimate (\ref{eq:UBEBS}) holds for every $x \in \T(t_0,u,\sigma)$ with $t_0\ge 0$, $u\in \U_m$ and 
$\sigma\in \S$.
 \begin{align} 
 \label{eq:UBEBS}
  |x(t)|\le \alpha_1(|x(t_0)|)+\alpha_2\left (\int_{t_0}^t\alpha(|u(s)|)\:ds\right )+c,\quad \forall t\ge t_0.
 \end{align}
\end{defin}
The following is the main result of this section.
\begin{teo}
  \label{UBEBSand0GASiffiiss}
  Let Assumption~\ref{ass:A} hold and let $\gamma$ be as in C\ref{item:bound}) of that assumption. Then, 
  \begin{enumerate}[a)]
  \item If system (\ref{eq:1}) is iISS w.r.t. $\S$ with iISS gain $\chi$, then (\ref{eq:1}) is 0-GUAS and UBEBS, both w.r.t. $\S$, with UBEBS gain $\chi$.\label{item:nec}
  \item If system (\ref{eq:1}) is 0-GUAS and UBEBS, both w.r.t. $\S$, with UBEBS gain $\alpha$ then (\ref{eq:1}) is iISS w.r.t. $\S$ with iISS gain $\chi = \max\{\alpha, \gamma\}$ and has the $\chi$-BEICS w.r.t. $\S$ property.\label{item:suff}
  \end{enumerate}
\end{teo}
Theorem~\ref{UBEBSand0GASiffiiss} contains a characterization of the iISS w.r.t. $\S$ property, namely that 0-GUAS + UBEBS $\Longleftrightarrow$ iISS (all w.r.t. $\S$). The statement of Theorem~\ref{UBEBSand0GASiffiiss} is split into parts \ref{item:nec}) and \ref{item:suff}) in order to keep track of the iISS gain. The proof of \ref{item:nec}) follows straightforwardly from Definitions~\ref{def:iiss}, \ref{def:0guas} and \ref{def:ubebs}. For establishing \ref{item:suff}) we require Lemmas~\ref{lem:0GAS} and~\ref{lem:0UBEBS} below, whose proofs are given in the Appendix. 
%
%
\begin{lema}
  \label{lem:0GAS} 
  Let Assumption \ref{ass:A} hold. Suppose that (\ref{eq:1}) is 0-GUAS w.r.t. $\S$ and let $\beta \in \KL$ characterize the 0-GUAS property, so that (\ref{eq:xest0guas}) is satisfied for the 0-input system. Let $\chi\in \Ki$ be such that $\chi(r)\ge \gamma(r)$ for all $r\ge 0$, with $\gamma$ as in C\ref{item:bound}) in Assumption \ref{ass:A}.
  Then, for every $r>0$ and every $\eta>0$ there exist $L=L(r) > 0$ and $\kappa = \kappa(r,\eta)$ such that the following holds:
if $x \in \T(t_0,u,\sigma)$, with $t_0\ge 0$, $u\in \U_m$ and $\sigma\in \S$, and $|x(t)|\le r$ for all $t\ge t_0$, then 
\begin{align}
|x(t)|&\le  \beta(|x(t_0)|,t-t_0)+ \nonumber \\
 &\quad \left [\eta (t-t_0)+\kappa \int_{t_0}^{t}\chi(|u(\tau)|)\:d\tau\right ]e^{L(t-t_0)}\quad\forall t\ge t_0.
\label{eq:estima}
 \end{align}
\end{lema}
Loosely speaking, Lemma~\ref{lem:0GAS} gives an estimate of how big the magnitude of the state can result depending on time and input energy, the latter in relation to the gain $\chi\in\Ki$, where the relative weights of the time- and energy-dependent terms can be modified. The estimate (\ref{eq:estima}) is useful only for small values of $t-t_0$, because $|x(t)| \le r$ for all $t\ge t_0$ is already assumed. 

Lemma~\ref{lem:0UBEBS} below shows that for a 0-GUAS system, UBEBS in Definition~\ref{def:ubebs} could be equivalently defined setting $c= 0$ in (\ref{eq:UBEBS}). The proof of this fact differs from the corresponding proof in Lemma~2.1 of \cite{angson_dc00} because a converse Lyapunov theorem cannot be invoked in the current setting.
\begin{lema}
  \label{lem:0UBEBS} 
  Consider system (\ref{eq:1}) and a set $\S$ of switching signals. Let Assumption~\ref{ass:A} hold, and let $\gamma$ be as in C\ref{item:bound}) of that assumption. If (\ref{eq:1}) is 0-GUAS and UBEBS, both w.r.t. $\S$, with UBEBS gain $\alpha$, then there exist $\tilde\alpha_1, \tilde\alpha_2\in \K$  for which the estimate (\ref{eq:0UBEBS}) holds with $\chi=\max\{\alpha,\gamma\}$ for every $x \in \T(t_0,u,\sigma)$ with $t_0\ge 0$, $u\in \U_m$ and $\sigma\in \S$.
 \begin{align} 
 \label{eq:0UBEBS}
  |x(t)|\le \tilde \alpha_1(|x(t_0)|)+\tilde \alpha_2\left (\int_{t_0}^t\chi(|u(s)|)\:ds\right )\quad \forall t\ge t_0.
 \end{align}
\end{lema}
 
\begin{IEEEproof}[Proof Theorem~\ref{UBEBSand0GASiffiiss}\ref{item:suff})]
%
Let $\chi=\max\{\alpha,\gamma\}$ and let $\|u\|= \|u\|_{\chi}$. Let $\tilde\alpha_1, \tilde\alpha_2$ be as in Lemma \ref{lem:0UBEBS}.  We will establish iISS with iISS gain $\chi$ w.r.t. $\S$ by following the items of Lemma~\ref{rem:iiss-charact}.

\ref{item:bound-iss}) Let $T>0$, $r>0$ and $s>0$. Let $x \in \T(t_0,u,\sigma)$ with $t_0\ge 0$, $u\in \U_m$ with $\|u\| \le s$ and $\sigma\in\S$, be such that $|x(t_0)|\le r$. From (\ref{eq:0UBEBS}), it follows that $|x(t)| \le \tilde \alpha_1(r)+\tilde \alpha_2(s) =: C$ for all $t\ge t_0$ because being bounded, $x$ cannot cease to exist. This establishes the forward completeness of (\ref{eq:1}) w.r.t. $\S$ and item~\ref{item:bound-iss}) of Lemma \ref{rem:iiss-charact}.

\ref{item:gs-iss}) Let $\epsilon > 0$. Let $\delta>0$ be such that $\tilde \alpha_1(\delta)+\tilde \alpha_2(\delta)<\epsilon$. Then, if $x \in \T(t_0,u,\sigma)$ with $t_0\ge 0$, $u\in \U_m$ with $\|u\| \le \delta$ and $\sigma\in\S$, and $|x(t_0)|\le \delta$, it follows, by using (\ref{eq:0UBEBS}), that $|x(t)|\le \tilde \alpha _1(\delta)+\tilde \alpha_2(\delta)<\epsilon$ for all $t\ge t_0$. This establishes item~\ref{item:gs-iss}) of Lemma~\ref{rem:iiss-charact}.

\ref{item:ua-iss}) Let $\nu\in\Ki$ be defined via $\nu(t) = 2 \tilde \alpha_2(t)$. Let $r\ge\epsilon>0$.  Let $x \in \T(t_0,u,\sigma)$ with $t_0\ge 0$, $u\in \U_m$ and $\sigma\in\S$, be such that $|x(t_0)| \le r$. Let $\phi\in\Ki$ be defined by $\phi(\cdot)=\tilde \alpha_2^{-1}\comp \tilde \alpha_1(\cdot)$. We distinguish two cases:
\begin{enumerate}[(a)]
\item $\|u\| \ge \phi(r)$,
\item $\|u\| < \phi(r)$.
\end{enumerate}
In case (a), from (\ref{eq:0UBEBS}) we have $|x(t)| \le \tilde \alpha_1(r) +\tilde\alpha_2(\|u\|) \le \tilde \alpha_1\comp\phi^{-1}(\|u\|) +\tilde\alpha_2(\|u\|) \le 2 \tilde \alpha_2(\|u\|)=\nu(\|u\|)$ for all $t\ge t_0$. Hence $|x(t)| \le \epsilon +\nu(\|u\|)$ for all $t\ge t_0$.

Next, consider case (b). From (\ref{eq:0UBEBS}), we have  $|x(t)|\le \tilde \alpha_1(r)+\tilde \alpha_2(\phi(r)):=\tilde r$
for all $t\ge t_0$. Let $\beta \in \KL$ characterize the 0-GUAS w.r.t. $\S$ property and let $L = L(\tilde r) > 0$ be given by Lemma \ref{lem:0GAS}. Let $\tilde \epsilon=\tilde \alpha_1^{-1}(\epsilon)$ and pick $\tilde T>0$ such that $\beta(\tilde r,\tilde T)< \tilde \epsilon/2$. Define $\eta=\frac{\tilde \epsilon}{4 \tilde T e^{L \tilde T}}$. Let $\kappa = \kappa(\tilde r,\eta) > 0$ be given by Lemma~\ref{lem:0GAS}. Pick $\delta>0$ such that $\kappa \delta e^{L \tilde T}< \tilde \epsilon/4$.  Define $N := \left\lceil \frac{\phi(r)}{\delta} \right\rceil$ and  $T:= N \tilde T$, where $\lceil s\rceil$ denotes the least integer not less than $s\in \R$.

For $i=0$ to $N$, let $t_i=t_0+i\tilde T$. Consider the intervals $I_i=[t_{i-1}, t_{i}]$, with $i=1,\ldots, N$. From the definition of $N$ and the fact that $\|u\|<\phi(r)$, there exists $j\le N-1$ for which $\int_{t_{j}}^{t_{j+1}} \chi(|u(s)|)\;ds<\delta$. Since $x\in \T(t_{j},u,\sigma)$ and $|x(t)|\le \tilde r$ for all $t\ge t_j$, and by using (\ref{eq:estima}),
  \begin{align*}
    |x(t_j+\tilde T)| &\le \beta(|x(t_j)|,\tilde T)+\left (\eta \tilde T+\kappa \int_{t_j}^{t_j+\tilde T} \chi(|u(s)|)\:ds\right ) e^{L\tilde T}\\
    &\le \beta(\tilde r,\tilde T)+(\eta \tilde T+\kappa \delta) e^{L\tilde T} <\tilde
\epsilon/2 + \tilde\epsilon/4 + \tilde\epsilon/4 = \tilde\epsilon.
  \end{align*}
Therefore, using (\ref{eq:0UBEBS}) with $t_0$ replaced by $t_j+\tilde{T}$, 
$$|x(t)|\le \tilde \alpha_1(\tilde \epsilon)+\tilde \alpha_2(\|u\|) \le \epsilon + \nu(\|u\|)\quad \forall t\ge t_0 + T$$ 
because $t_0+T\ge t_j+\tilde T$, which shows that item~\ref{item:ua-iss}) of Lemma~\ref{rem:iiss-charact} also is satisfied.

Finally, that (\ref{eq:1}) has the $\chi$-BEICS property w.r.t. $\S$ follows from Proposition \ref{prop:iISSimpliesBEICS}. 
\end{IEEEproof}

\section{Dissipativity and iISS}
\label{sec:dissipativity-iiss}

In this section, we show that other characterizations of iISS valid for time-invariant and non-switched systems, e.g. those based on dissipativity and weak detectability (see \cite{angson_tac00,jayrya_tac10}), only give sufficient conditions in the current setting. 
We next consider system (\ref{eq:1}) with an output of the form
\begin{align}\label{eq:2}
 y=h(t,x,u,\sigma),
\end{align}
where $h:\R_{\ge 0}\times
\R^n\times \R^m \times \Gamma\to \R^p$ is continuous in the second and third variables and Lebesgue measurable in the first one. We also assume that $h_0(t,\xi,i)\equiv h(t,\xi,0,i)$ is essentially bounded on $\R_{\ge 0} \times K \times \Gamma$, for every compact subset $K \subset \R^n$ such that $0\notin K$.

Definition~\ref{def:hOD} below extends the dissipativity notion to our setting. Definition~\ref{def:0iope} replaces the notion of weak detectability by a suitable extension in terms of persistence of excitation (see the subsequent Remark~\ref{rem:w0sdimplies0iope}).
\begin{defin} 
  \label{def:hOD}
  Let $\S$ be a set of switching signals. System (\ref{eq:1}) with output (\ref{eq:2}) is called $h$-output dissipative ($h$-OD) w.r.t. $\S$ if there exist a function $V : \R_{\ge 0} \times \R^n \to \R_{\ge 0}$ (the storage function) and a function $\alpha\in\K$ (the supply function) such that \ref{item:Vphi12}) and \ref{item:dissip}) below hold:
  \begin{enumerate}[a)] 
   \item \label{item:Vphi12}There exist $\phi_1$ and $\phi_2\in\Ki$ so that
  \begin{align}
    \label{eq:Vphi12}
    \phi_1(|\xi|) \le V(t,\xi) \le \phi_2(|\xi|) \quad \forall t\ge 0, \forall \xi\in\R^n.
    \end{align}
  \item \label{item:dissip}There exists a continuous and positive definite function $\alpha_3$ such that  
  for every $x\in \T(t_0,u,\sigma)$ with $t_0\ge 0$, $u\in \U_m$ and $\sigma\in \S$,
 \begin{multline}
    V(t,x(t)) \le V(t_0,x(t_0))
   -\int_{t_0}^t \alpha_3(|y(\tau)|)\:d\tau\\
    \label{eq:hOD}
   + \int_{t_0}^t \alpha(|u(\tau)|) d\tau,
    \quad\forall t\ge t_0.
  \end{multline}
\end{enumerate}
System (\ref{eq:1}) is called zero-output dissipative (0-OD) w.r.t. $\S$ if it is $h$-output dissipative w.r.t. $\S$ for the output $y=0$, i.e. the output map $h$ is $h=0$.
Note that $h$-OD w.r.t. $\S$ implies 0-OD w.r.t. $\S$. 
\end{defin}
\begin{remark}
  \label{rem:Vnotcont}
  We highlight the fact that the storage function $V$ need not even be continuous. However, continuity at $(t,0) \in \R_{\ge 0} \times \R^n$ for every $t\ge 0$ follows from item~\ref{item:Vphi12}) of Definition~\ref{def:hOD}.\mer
\end{remark}
\begin{defin} 
  \label{def:0iope}
  We say that the pair $(h,f)$ is zero-input output-persistently exciting (output-PE) w.r.t. $\S$ if for every $0<\varepsilon\le 1$ there exist $T=T(\varepsilon)>0$ and  $r=r(\varepsilon)>0$ 
  such that for every $x \in \T(t_0,\mathbf{0},\sigma)$ with $t_0\ge 0$ and $\sigma\in \S$ and every $t\ge t_0$ the following implication holds
  \begin{multline} 
    \label{OPE}
    \varepsilon \le |x(\tau)| \le \frac{1}{\varepsilon}, \quad \forall \tau\in [t,t+T]\quad \Longrightarrow \\
    \int_{t}^{t+T} |h_0(\tau,x(\tau),\sigma(\tau))|^2\:d\tau \ge r. 
  \end{multline}
\end{defin}
\begin{remark} 
  \label{rem:w0sdimplies0iope}
  It can be easily proved that in the case of a non-switched time-invariant system $\dot x=f(x,u)$ with outputs $y=h(x,u)$, weak zero-detectability as defined in \cite{angson_tac00} is equivalent to the zero-input output-PE property of the pair $(h,f)$.
 \end{remark}
The following lemma will be used in the proof of Theorem \ref{thm:suffcondsiISS}. It can be proved using Corollary 1 in \cite{leetan_tac15}. For the reader's convenience, we provide a proof in the Appendix.
\begin{lema} \label{lem:ope}
Suppose that $(h,f)$ is zero-input output-PE w.r.t. $\S$. Then, for any continuous positive definite function $\alpha$, the pair $(\hat{h},f)$, with $\hat{h}=\alpha(|h|)$, is zero-input output-PE w.r.t. $\S$.
\end{lema}
Theorem~\ref{thm:suffcondsiISS} below provides sufficient conditions for iISS. The fact that these conditions are only sufficient and not necessary will be established later in this section. 
\begin{teo}
  \label{thm:suffcondsiISS}
  Let Assumption \ref{ass:A} hold and let $\gamma$ satisfy C\ref{item:bound}) in Assumption~\ref{ass:A}. Then, (\ref{eq:1}) is iISS with iISS gain $\chi=\max\{\alpha,\gamma\}$ and has the $\chi$-BEICS property, both w.r.t. $\S$, if either of the following conditions holds:
  \begin{enumerate}[a)]
  \item System (\ref{eq:1}) is 0-GUAS and 0-OD with supply function
    $\alpha \in \K$, both w.r.t. $\S$.\label{item:00ODimplyiISS}
  \item There exists an output (\ref{eq:2}) for which (\ref{eq:1}) is $h$-OD w.r.t. $\S$ with supply function $\alpha$ and $(h,f)$ is zero-input output-PE w.r.t. $\S$. \label{item:hOD-0OPEimpliesiISS}
  \end{enumerate}
\end{teo}
\begin{IEEEproof}
  \ref{item:00ODimplyiISS}). From the definitions of UBEBS and 0-OD (Definitions~\ref{def:ubebs} and~\ref{def:hOD}), it straightforwardly follows that if system (\ref{eq:1}) is 0-OD w.r.t. $\S$ with supply function $\alpha\in \K$ then it is UBEBS w.r.t. $\S$ with UBEBS gain $\alpha$. The proof of \ref{item:00ODimplyiISS}) then follows by application of Theorem~\ref{UBEBSand0GASiffiiss}. 

  \ref{item:hOD-0OPEimpliesiISS}) We first prove that (\ref{eq:1}) is 0-GUAS w.r.t. $\S$. Let $\Phi$ be the set of all the pairs $(x,\sigma)$ with $x\in \T(t_0,\mathbf{0},\sigma)$ with $t_0\ge 0$ and $\sigma\in \S$. We will show that $\Phi$ satisfies the hypotheses of Theorem 1 in \cite{leejia_tac08}, and that in consequence $\Phi$ is uniformly globally asymptotically stable (in the sense of \cite{leejia_tac08}), which, in turn, implies that (\ref{eq:1}) is  0-GUAS w.r.t. $\S$.
In fact, from the $h$-OD condition and by using standard techniques of stability theory it follows that $\Phi$ is uniformly globally stable (in the sense of \cite{leejia_tac08}). Let $\alpha_3$ be the continuous and positive definite function appearing in (\ref{eq:hOD}), and let $\hat{h}=\sqrt{\alpha_3(|h|)}$. Since $(h,f)$ is zero-input output-PE w.r.t. $\S$, by Lemma \ref{lem:ope} it follows that  $(\hat{h},f)$ also is zero-input output-PE w.r.t. $\S$. Then $(\hat{h}_0,f_0)$, with $f_0(t,\xi,i)\equiv f(t,\xi,0,i)$, is output-PE w.r.t. $\Phi$ (in the sense of \cite{leejia_tac08}). Finally, from the $h$-OD condition it also follows that condition (H1) in \cite{leejia_tac08} is satisfied by $\hat{h}_0$ and $\Phi$ (see Remark 8 in that paper).
Since (\ref{eq:1}) is 0-GUAS and also 0-OD (because $h$-OD implies 0-OD), both w.r.t. $\S$, then application of part~\ref{item:00ODimplyiISS}), establishes that (\ref{eq:1}) is iISS with iISS gain $\chi$ and has the $\chi$-BEICS property, both w.r.t. $\S$.
\end{IEEEproof}


 %
\begin{remark} 
  \label{rem:extention} 
  Theorem~\ref{thm:suffcondsiISS}\ref{item:00ODimplyiISS}) contains the main result of \cite{jayrya_tac10} (Theorem 3.1) as a particular case, since our assumptions are weaker than those in \cite{jayrya_tac10}. Again, we remark that the corresponding proof in \cite{jayrya_tac10} does not apply in the current setting since that proof is based on the existence of a continuously differentiable Lyapunov function for the 0-input system. Such a Lyapunov function need not exist in the current setting, even for a time-varying system without switching.\mer
\end{remark}
%
%
%
%

Theorem~\ref{thm:suffcondsiISS}\ref{item:00ODimplyiISS}) and \ref{thm:suffcondsiISS}\ref{item:hOD-0OPEimpliesiISS}) are extensions of, respectively, the implications 4 $\Rightarrow$ 1 and 3 $\Rightarrow$ 1 in Theorem 1 in \cite{angson_tac00} to time-varying both switched and non-switched systems. We note that the corresponding proofs in \cite{angson_tac00} cannot be directly adapted since they heavily rely on converse Lyapunov theorems which do not exist in the current setting. In Theorem 1 in \cite{angson_tac00} it is shown that the converse of those implications also holds. Unfortunately, the converse of Theorem~\ref{thm:suffcondsiISS}\ref{item:00ODimplyiISS}) or \ref{item:hOD-0OPEimpliesiISS}) does not hold in our case. To prove the latter assertion, it suffices to show that there exists a system which is iISS w.r.t. some family of switching signals $\S$ and which is not 0-OD w.r.t. $\S$. 
\begin{prop} 
  \label{prop:onlysuff}
  There exist a system (\ref{eq:1}) and a set of switching signals $\S$ such that the system is iISS w.r.t. $\S$ but not 0-OD w.r.t. $\S$.
\end{prop}
To prove this proposition, we require some additional definitions and results.  
Given switching signals $\sigma_1,\ldots,\sigma_k$ and a sequence of times $0< t_1<\ldots<t_{k-1}$, the concatenation of them at times $t_1,\ldots,t_{k-1}$ is the switching signal 
  $$ \sigma_1\sharp_{t_1}\sigma_2\sharp_{t_2}\cdots \sharp_{t_{k-1}}\sigma_k(s)=\left \lbrace\begin{array}{clc} \sigma_1(s) & \makebox{if}& s<t_1 \\
                                  \sigma_2(s)& \makebox{if} & t_1\le s < t_2 \\
                                  \vdots & \vdots & \vdots \\
                                  \sigma_k(s) &  \makebox{if} & t_{k-1}\le s
                                 \end{array} \right . $$
Let $\S^\sharp_k$ denote the set of all the switching signals obtained by concatenating $k$ switching signals in $\S$ and let $\S^{\sharp}=
\cup_{k\ge 2} \S^{\sharp}_k$. Note that 
$\S\subset \S^{\sharp}_k \subset \S^{\sharp}$ for any $k\ge 2$, since $\sigma=\sigma \sharp_t \sigma$ for every $\sigma\in \S$ and every $t> 0$.

\begin{lema} 
  \label{lem:concatenation} 
  If (\ref{eq:1}) is 0-OD w.r.t. $\S$ then (\ref{eq:1}) is 0-OD w.r.t. $\S^{\sharp}$.
\end{lema}
\begin{IEEEproof}
  Let $V$ be a storage function as per Definition~\ref{def:hOD}, corresponding to the 0-OD w.r.t. $\S$ property. In order to show that (\ref{eq:1}) is 0-OD w.r.t. $\S^{\sharp}$ it suffices to show that the estimate (\ref{eq:hOD}), with $y=0$, holds for every $x\in \T(t_0,u,\sigma)$ with $t_0\ge 0$, $u\in \U_m$ and $\sigma\in \S^{\sharp}$. By induction in $k$, we will prove for all $k\ge 2$ that the estimate (\ref{eq:hOD}), with $y=0$, holds for every $x\in \T(t_0,u,\sigma)$ with $t_0\ge 0$, $u\in \U_m$ and $\sigma\in \S^{\sharp}_k$.

  {\em Case $k=2$}. Let $\sigma=\sigma_1\sharp_{t_1}\sigma_2$ with $\sigma_i\in \S$ for $i=1,2$ and $t_1>0$. Let $x \in \T(t_0,u,\sigma)$ with $t_0\ge 0$ and $u\in \U_m$. If   $t_1\le t_0$ then by well-known results on differential equations and causality there exists $x^*\in \T(t_0,u,\sigma_2)$ such that $x(t)=x^*(t)$ for all $t\ge t_0$. Since (\ref{eq:hOD}) with $y=0$ holds for $x^*$, it also holds for $x$. If $t_0<t_1$, then by causality and well-known results on differential equations there exists $x_1\in \T(t_0,u,\sigma_1)$ and $x_2\in \T(t_1,u,\sigma_2)$ such that $x(t)=x_1(t)$ for all $t\in [t_0,t_1]$ and $x(t)=x_2(t)$ for all $t\in [t_1,\infty)$, whence $x_1(t_1)=x_2(t_1)$. Then, for all $t\in [t_0,t_1]$ we have
  \begin{align*}
    V(t,x(t))=V(t,x_1(t))&\le V(t_0,x_1(t_0))+\int_{t_0}^{t}\chi(|u(s)|)\;ds \\
                         &= V(t_0,x(t_0))+\int_{t_0}^{t}\chi(|u(s)|)\;ds 
  \end{align*}
  and for all $t\ge t_1$,
  \begin{align*}
    V(t,x(t))&\le V(t_1,x_2(t_1))+\int_{t_1}^{t}\chi(|u(s)|)\;ds \\
             &\le V(t_0,x(t_0))+\int_{t_0}^{t_1}\chi(|u(s)|)\;ds + \int_{t_1}^{t}\chi(|u(s)|)\;ds \\
             &= V(t_0,x(t_0))+\int_{t_0}^{t}\chi(|u(s)|)\;ds.
  \end{align*}
  {\em Recursive step}. Suppose that (\ref{eq:hOD}) with $y=0$ holds for every $x\in \T(t_0,u,\sigma)$ with $\sigma \in \S^{\sharp}_k$. Let $\sigma\in \S^{\sharp}_{k+1}$. Then, there exist $\sigma_1,\ldots,\sigma_{k+1}$ in $\S$ and a sequence of times $0< t_1<\ldots<t_{k}$ such that $\sigma=\sigma_1\sharp_{t_1}\sigma_2\sharp_{t_2}\cdots    \sharp_{t_{k}}\sigma_{k+1}$. Let $\tilde\sigma=\sigma_1\sharp_{t_1}\sigma_2\sharp_{t_2}\cdots
  \sharp_{t_{k-1}}\sigma_k \in \S^{\sharp}_k$. Then $\sigma=\tilde \sigma \sharp_{t_{k}}\sigma_{k+1}$. Taking into account that (\ref{eq:hOD}) with $y=0$ holds for all the solutions $x$ of (\ref{eq:1}) corresponding to switching signals in $\S^{\sharp}_k\supset\S$, and by using the same arguments as in the case $k=2$, we can conclude that (\ref{eq:hOD}) with $y=0$ holds for all solutions $x$ of (\ref{eq:1}) corresponding to switching signals in $\S^{\sharp}_{k+1}$.
\end{IEEEproof}

\begin{IEEEproof}[Proof of Proposition~\ref{prop:onlysuff}]
Consider (\ref{eq:1}) with $f(t,\xi,\mu,i)=A_i\xi+b_i\mu$, $i=1,2$, where 
\begin{align*}
  A_1 &= \left[
        \begin{array}{rr}
          -1 & -100 \\
          10 & -1
        \end{array}\right],
 &A_2 &= A_1',
 &b_1 &= b_2 =
        \begin{bmatrix}
          1\\ 0
        \end{bmatrix}.
\end{align*}
Let $\S=\{\sigma_1,\sigma_2\}$, where $\sigma_i(t)=i$ for all $t\ge 0$. We note that both $A_1$ and $A_2$ are Hurwitz, and in consequence each subsystem, i.e. each of the two systems $\dot x = f(t,x,u,i)$, with $i=1,2$, is iISS. Then, (\ref{eq:1}) is iISS w.r.t. $\S$. We claim that (\ref{eq:1}) is not 0-OD w.r.t. $\S$. 

For a contradiction, suppose that (\ref{eq:1}) is 0-OD w.r.t. $\S$. By Lemma~\ref{lem:concatenation}, then (\ref{eq:1}) is 0-OD w.r.t. $\S^{\sharp}$. Therefore, there exists a storage function $V$ verifying (\ref{eq:Vphi12}), with $\phi_1,\phi_2\in \Ki$, and such that for some $\chi\in \Ki$, (\ref{eq:hOD}) holds with $y=0$ for every $x\in \T(t_0,u,\sigma)$, with $t_0\ge 0$, $u\in \U_1$ and $\sigma\in \S^\sharp$. It then follows that every solution $x$ of $\dot{x}=A_{\sigma}x$ with $\sigma \in \S^{\sharp}$ must verify $$ V(t,x(t))\le V(t_0,x(t_0))\quad \forall t\ge t_0.$$
In particular, for every $r>0$ there exists $c(r)\ge 0$ such that for each switching signal $\sigma \in \S^{\sharp}$ and each $\xi_0\in \R^2$ with $|\xi_0|\le r$, the unique solution $x(t,\xi_0,\sigma)$ of $\dot{x}=A_{\sigma}x$, $x(0)=\xi_0$ satisfies $|x(t,\xi_0,\sigma)|\le c(r)$ for all $t\ge 0$.

Also, for each initial condition $\xi_0\neq 0$, there is a switching signal $\sigma_{\xi_0}:\R_{\ge 0}\to \{1,2\}$ ($\sigma_{\xi_0}$ does not necessarily satisfy $\sigma_{\xi_0} \in \S^\sharp$) such that the unique solution $x(t,\xi_0,\sigma_{\xi_0})$ of $\dot{x}=A_{\sigma_0}x$, $x(0)=\xi_0$ satisfies $|x(t,\xi_0,\sigma_{\xi_0})|\to \infty$ (see Example 2 in \cite{decbra_pieee00}).

Pick any $\xi_0\neq0$ and let $r=|\xi_0|$. Then there exists $T>0$ such that $|x(T,\xi_0,\sigma_{\xi_0})|>c(r)$. From the definition of $\S^{\sharp}$ and the fact that any switching signal has a finite number of discontinuities in every bounded interval, it can be easily seen that the switching signal $\tilde \sigma=\sigma_{\xi_0}\sharp_T \sigma_1$ belongs to $\S^{\sharp}$. By causality, we have that $|x(T,\xi_0,\sigma_{\xi_0})|=|x(T,\xi_0,\tilde \sigma)|\le c(r)$. Since we have arrived to a contradiction, it follows that system (\ref{eq:1}) is not 0-OD w.r.t. $\S$. 
\end{IEEEproof}
\section{Example}
\label{sec:examples}

We provide an example to illustrate the application of Theorem~\ref{thm:suffcondsiISS}.
  Consider the ideal switched model of the semi-quasi-Z-source inverter \cite{caojia_tpe11,haimid_cdc13}, connected to a nonlinear time-varying resistive load and where $u$ represents the input voltage:
  \begin{align}
  \label{eq:ex1}  \dot x &= f(t,x,u,\sigma) = \tilde A_{\sigma} x - e_4 \tilde g_{\sigma}(t,e_4'x) + b_{\sigma} u,\\
    e_4 &= [0\ 0\ 0\ 1]',\quad
    P = \diag(L_1,L_2,C_1,C_2)\notag\\
    \tilde A_1 &= P^{-1}\left[
      \begin{smallmatrix}
        0 & 0 & 0 & 0\\
        0 & 0 & 1 & 1\\
        0 & -1 & 0 & 0\\
        0 & -1 & 0 & 0
      \end{smallmatrix}\right],\quad
    \tilde A_2 = P^{-1}\left[
      \begin{smallmatrix}
        0 & 0 & -1 & 0\\
        0 & 0 &  0 & 1\\
        1 & 0 &  0 & 0\\
        0 & -1 & 0 & 0
      \end{smallmatrix}\right],\notag\\
    b_1 &= P^{-1} [1\ 0\ 0\ 0]',\quad b_2 = P^{-1} [0\ 1\ 0\ 0]',\notag\\
    \label{eq:load}
    \tilde g_i(t,v) &= a_i(t)\sat(v/r_i(t)),
  \end{align}
  where $\sat$ is the unitary saturation function ($\sat(v) = v$ if $|v| \le 1$ and $\sat(v) = \sign(v)$ otherwise), and $a_i(t) \in [a_{\min},a_{\max}]$, $a_{\max} \ge a_{\min} > 0$, and $r_i(t) \in [r_{\min},r_{\max}]$, $r_{\max}\ge r_{\min} > 0$, 
  for all $t\ge 0$ and for $i=1,2$. The positive constants $L_1,L_2,C_1,C_2$ represent the inverter 
  inductance and capacitance values. It is clear that this system verifies Assumption \ref{ass:A} and, in particular, C\ref{item:bound}) is satisfied, e.g., with $\gamma(s)=s$ and $N(|\xi|) = \max\{\|A_1\|,\|A_2\|\} |\xi| + a_{\max} + \max\{1,\|b_1\|,\|b_2\|\}$, with $\|A_i\|$ the matrix norm induced by the Euclidean vector norm. 
  
  Irrespective of the load function $\tilde g_i$, stability of this inverter model 
  can only be ensured by constantly switching between $\sigma(t)=1$ (mode 1) and $\sigma(t) = 2$ (mode 2), and imposing additional restrictions on the time spent 
  in mode 2 \cite{denhai_auto16}. Let $\S$ denote the set of switching signals $\sigma : \R_{\ge 0} \to \{1,2\}$ where 
  each mode has minimum ($d_{\min}$) and maximum ($d_{\max}$) dwell times satisfying 
  $0 < d_{\min} < d_{\max} < \pi \sqrt{L_1C_1}$. 
  
  In order to show that the system is 0-GUAS and 0-OD w.r.t. $\S$, 
  we consider the time-invariant positive definite quadratic function 
  $V(t,x) = \bar V(x) = \frac{1}{2}x'Px$. Such a function satisfies (\ref{eq:Vphi12}), with $\phi_1(s) = \lambda_{\min}s^2$ and $\phi_2(s)=\lambda_{\max}s^2$ with $\lambda_{\min},\lambda_{\max}$ the minimum and maximum eigenvalues of $P/2$, and $\dot V_i$, its derivative along the trajectories of the $i$th-subsystem, is
  \begin{multline*}
    \dot{V}_i(t,\xi)=
    \xi'P\tilde A_i \xi - \xi' C_2 e_4 \tilde g_i(t,e_4'\xi)+\xi'Pb_i\mu\\
    = -\underbrace{C_2 (e_4'\xi) a_i(t)\sat(e_4'\xi/r_i(t))}_{\eta_i(t,\xi)}+\xi'Pb_i\mu.
  \end{multline*}
  Note that for $i=1,2$, $P\tilde A_i$ is skew-symmetric and hence $\xi'P\tilde A_i \xi = 0$ for all $\xi\in\R^4$ and that $\eta_i$ is a nonnegative function because $C_2 > 0$, $a_i(t) > 0$, $r_i(t) > 0$, and $v\,\sat(v/r_i(t)) \ge 0$ for all $v\in\R$. 
  To show that this system is 0-GUAS w.r.t. $\S$, we employ Theorem 3.1 of \cite{manhai_tac17}. This requires decomposing the zero-input system equations into a ``nominal'' 0-GUAS part $\hat f$ and a ``perturbation'' part $g$, as follows:
  \begin{align*}
   \dot x = f(t,x,0,\sigma) = \underbrace{(\tilde A_{\sigma} - K e_4 e_4')x}_{\hat f(t,x,\sigma)} \underbrace{- e_4 (\tilde g_{\sigma}(t,e_4'x) - K e_4'x)}_{g(t,x,\sigma)},
  \end{align*}
  where $K>0$ is an arbitrary constant. Under this decomposition, Assumption 1 of Theorem 3.1 of \cite{manhai_tac17} is satisfied because the switched linear system $\dot x = \hat f(t,x,\sigma)$ is 0-GUAS w.r.t. $\S$, as established in \cite{haimid_cdc13}. The function $V$ and $\eta_i$ as above satisfy Assumption 2, and the functions $\hat f$ and $g$ satisfy the boundedness condition of Assumption 3 of Theorem 3.1 of \cite{manhai_tac17}. Finally, the functions $\eta(t,\xi,i) = \eta_i(t,\xi)$ and $g$ satisfy condition (C) of the latter theorem, and hence the zero-input system is 0-GUAS w.r.t. $\S$. 
Next, since $\eta_i \ge 0$, then 
  \begin{align*}
     \dot{V}_i(t,\xi)\le \xi'Pb_i u\le \kappa \sqrt{V} |u|,
  \end{align*}
with $\kappa = 1/\sqrt{\lambda_{\min}}$. Then, using a comparison lemma for differential equations we have that for every solution $x$ of (\ref{eq:ex1})
corresponding to an input $u$ and a switching signal $\sigma\in \S$, the following holds
\begin{align*}
 \sqrt{V(t,x(t))}\le \sqrt{V(t_0,x(t_0))}+\frac{\kappa}{2} \int_{t_0}^t|u(s)|\:ds.
\end{align*}
In consequence, system (\ref{eq:ex1}) is 0-OD w.r.t. $\S$ if we consider $\sqrt{V}$ as the storage function. By Theorem~\ref{thm:suffcondsiISS}\ref{item:00ODimplyiISS}), system \eqref{eq:ex1} is iISS w.r.t. $\S$ with iISS gain $\chi(s)=s$. It also has the $\chi$-BEICS property. Thus, for $x\in \T(t_0,u,\sigma)$ with $u \in L^1(\R_{\ge 0})$ and $\sigma\in \S$ it follows that $x(t)\to 0$ as $t\to \infty$.

We remark that system \eqref{eq:ex1} is not ISS w.r.t. $\S$. Indeed, the second simulation example in Section 2.5 of \cite{denhai_auto16} corresponds to the considered system with a load of the form \eqref{eq:load} for constant and positive $a_i$ and $r_i$, for $i=1,2$, and a switching signal contained in the considered set $\S$. This simulation shows that the state is divergent for a bounded input $u$, and hence the system cannot be ISS w.r.t. $\S$. Another interesting fact about this example is that no iISS common Lyapunov function exists, because none of the subsystems is 0-GUAS.

\section{Conclusions}
\label{sec:concl}

We have provided a characterization of integral input-to-state stability that is valid for switched and time-varying systems uniformly over arbitrary sets of switching signals. We have also shown that some natural extensions of the characterizations available for non-switched time-invariant systems become only sufficient conditions in the setting considered. Our proofs are novel in the sense that no converse Lyapunov theorems are required.  

\appendix
\subsection{Proof of Lemma \ref{lem:loc-lips}}
Conditions C\ref{item:fcont}) and C\ref{item:lips}) in Assumption \ref{ass:A} are obviously satisfied. We proceed to prove C\ref{item:bound}). The function
$$ \tilde \gamma(r):=\sup\{|f(t,\xi,\mu,i)|:\;t\ge 0, i\in \Gamma,\;|\xi|\le r,\;|\mu|\le r\}$$
is clearly nondecreasing, and finite for all $r\ge 0$ because of the assumptions of Lemma~\ref{lem:loc-lips}. In addition, if $\tilde{L}>0$ is a Lipschitz constant for $f(t,\cdot,\cdot,i)$ on the compact set $\{(\xi,\mu)\in \R^{n}\times \R^{m}:\;|\xi|\le 1,\;|\mu|\le 1\}$, then $\tilde{\gamma}(r)\le 2\tilde{L} r$ for all $0\le r\le 1$.
In consequence, there exists $\gamma\in \Ki$ such that $\gamma(r)\ge \tilde{\gamma}(r)$ for all $r\ge 0$ and such that $\gamma(r)=2\tilde{L} r$ for all $0\le r \le 1/2$. We note that $|f(t,\xi,\mu,i)|\le \gamma(|\xi|)+\gamma(|\mu|)\le N(|\xi|)[1+\gamma(|\mu|)]$, with $N(r)=\max\{1,\gamma(r)\}$.\hfill\QED
%
%
\subsection{Proof of Lemma \ref{lem:0GAS}}
\label{sec:proof-lemma-refl}

For any $p\in\N$ and $s>0$, we define $\bar{B}^p_s := \{\xi \in \R^p:\;|\xi|\le s\}$.\\
{\em Claim:} For every $r^*>0$ and $\eta>0$ there exists $\kappa=\kappa(r^*,\eta)>0$ such that for all $t\ge 0$, $\xi \in \bar{B}^n_{r^*}$, $\mu \in \R^m$ and $i\in \Gamma$,
\begin{align}\label{eq:inI}
 |f(t,\xi,\mu,i)-f(t,\xi,0,i)|\le \eta+\kappa \gamma(|\mu|).
\end{align}
From C\ref{item:fcont}) in Asumption \ref{ass:A} there exists $0<\delta<1$ such that for all $t\ge 0$, $i\in \Gamma$, and $(\xi,\mu)\in \bar{B}^n_{r^*}\times \bar{B}^m_{\delta}$, $|f(t,\xi,\mu,i)-f(t,\xi,0,i)|<\eta$. 
If $\xi\in \bar{B}^n_{r^*}$ and $|\mu|\ge \delta$, using C\ref{item:bound}) it follows that $|f(t,\xi,\mu,i)-f(t,\xi,0,i)| \le |f(t,\xi,\mu,i)| + |f(t,\xi,0,i)| \le 2N(|\xi|) + N(|\xi|) \gamma(|\mu|) \le 2N(r^*) + N(r^*)\gamma(|\mu|)$ and hence $|f(t,\xi,\mu,i)-f(t,\xi,0,i)|/\gamma(|\mu|)\le N(r^*)[2/\gamma(\delta)+1]=:\kappa$. In consequence
$$  |f(t,\xi,\mu,i)-f(t,\xi,0,i)|\le \kappa \gamma(|\mu|)\quad \forall \xi\in \bar{B}_{r^*}^n, |\mu|\ge \delta.$$
Combining the inequalities obtained, the claim is established.

Next, let $r>0$ and $r^*=\beta(r,0)\ge r$. Let $L=L(r) > 0$ be any Lipschitz constant for $f(t,\cdot,0,i)$ on the compact set $\bar{B}^n_{r^*}$ valid for every $t\ge 0$ and every $i\in \Gamma$.  Let $x\in \T(t_0,u,\sigma)$ with $t_0\ge 0$, $u\in \U_m$ and $\sigma\in \S$ be such that $|x(t)|\le r$ for all $t\ge t_0$.  Let $x_0\in \T(t_0,\mathbf{0},\sigma)$ be such that $x_0(t_0) = x(t_0)$. Then, both $x$ and $x_0$ evolve in $\bar{B}^n_{r^*}$ for all $t\ge t_0$.
Let $t\ge t_0$. For all $t_0\le \tau\le t$, we have
\begin{align*}
  \lefteqn{|x(\tau) - x_0(\tau)|}\hspace{5mm}\\
&\le \int_{t_0}^{\tau} |f(s,x(s),u(s),\sigma(s)) - f(s,x_0(s),0,\sigma(s))| ds \\
&\le \int_{t_0}^{\tau} |f(s,x(s),u(s),\sigma(s)) - f(s,x(s),0,\sigma(s))| ds\\
&\hspace{5mm} + \int_{t_0}^{\tau} |f(s,x(s),0,\sigma(s)) - f(s,x_0(s),0,\sigma(s))| ds\\
&\le \int_{t_0}^{\tau} [\eta+\kappa \gamma(|u(s)|)] ds + \int_{t_0}^{\tau} L |x(s) - x_0(s)| ds \\
&\le \eta(t-t_0)+\kappa \int_{t_0}^{t} \chi(|u(s)|) ds + \int_{t_0}^{\tau} L |x(s) - x_0(s)| ds.
\end{align*}
Using Gronwall's inequality, it follows that
\begin{align*}
  |x(t) - x_0(t)| &\le \left [\eta(t-t_0)+ \kappa \int_{t_0}^{t} \chi(|u(s)|) ds \right ]  e^{L(t-t_0)}\quad \forall t\ge t_0.
\end{align*} 
The lemma is then established from $|x(t)| \le |x_0(t)| + |x(t) - x_0(t)|$ and recalling the estimate (\ref{eq:xest0guas}) for $x_0(t)$.\hfill\QED

\subsection{Proof of Lemma~\ref{lem:0UBEBS}}
\label{sec:pf-lem-ubebsc0}

  Let $\alpha_1$, $\alpha_2$, $\alpha$ and $c$ be as in the estimate
  (\ref{eq:UBEBS}) and let $\chi=\max\{\alpha,\gamma\}$.  For
  $r\ge 0$ define
$$ \tilde \alpha(r):=\sup_{x\in \mathcal{T}(t_0,u,\sigma),\; t\ge t_0\ge 0,\; \|u\|\le r,\; \sigma\in \S, \;|x(t_0)|\le r} |x(t)| $$
where $\|u\|:=\|u\|_{\chi}$. From this definition, it follows that $\tilde\alpha$ is nondecreasing and from (\ref{eq:UBEBS}) that it is finite for all $r\ge 0$.
Next, we show that $\lim_{r\to 0^+}\tilde \alpha(r)=0$. Let $r^*=\alpha_1(1)+\alpha_2(1)+c$, $\beta\in \KL$ be the function which characterizes the 0-GUAS w.r.t. $\S$ property and $L=L(r^*)>0$ be given by Lemma \ref{lem:0GAS}. 
Let $\varepsilon>0$ be arbitrary. Pick $0<\delta_1<1$ such that $\delta_1 \le \beta(\delta_1,0)<\varepsilon/2$, and $T>0$ such that $\beta(\delta_1,T)<\delta_1/2$. Define $\eta=\frac{\delta_1}{4 T e^{LT}}$ and let $\kappa=\kappa(r^*,\eta)>0$ be given by Lemma \ref{lem:0GAS}. Last, pick $0<\delta_2<1$ such that $\kappa \delta_2 e^{LT}<\delta_1/4$. Then, for every $x\in \T(t_0,u,\sigma)$, with $t_0\ge 0$, $u\in \U_m$
with $\|u\|\le \delta_2$, $\sigma\in \S$ and $|x(t_0)|\le \delta_1$ we
claim that $|x(t)|<\varepsilon$ for all $t\ge t_0$. In fact, for all
$t\in [t_0,t_0+T]$, we have from Lemma \ref{lem:0GAS} that
$|x(t)|\le \beta(|x(t_0)|,t-t_0)+ (\eta(t-t_0)+\kappa \|u\|)e^{L(t-t_0)}\le
\beta(\delta_1,0)+(\eta T+\kappa \delta_2) e^{LT}<\varepsilon$
and that
$|x(t_0+T)|\le \beta(\delta_1,T)+(\eta T+ \kappa \delta_2)
e^{LT}<\delta_1$.
Since $x\in \T(t_1,u,\sigma)$, with $t_1=t_0+T$, and
$|x(t_1)|< \delta_1$, then $|x(t)|<\varepsilon$ for all
$t\in [t_1,t_1+T]$ and $|x(t_1+T)|<\delta_1$. Therefore, by using an
inductive argument we can prove that $|x(t)|<\varepsilon$ for all
$t\in [t_n,t_{n}+T]$, where $t_n=t_0+nT$, and that
$|x(t_n+T)|<\delta_1$. In consequence, $|x(t)|<\varepsilon$ for all
$t\ge t_0$ as we claim.
Thus, if $\delta=\min\{\delta_1,\delta_2\}$, for all
$x\in \T(t_0,u,\sigma)$, with $t_0\ge 0$, $u\in \U_m$ with
$\|u\|\le \delta$, $\sigma\in \S$ and $|x(t_0)|\le \delta$, we have
$|x(t)|\le \varepsilon$ for all $t\ge t_0$. Therefore,
$\tilde \alpha(r)\le \tilde \alpha(\delta)<\varepsilon$ for all
$0<r<\delta$ and $\lim_{r\to 0^+}\tilde \alpha(r)=0$.
 
Since $\tilde \alpha$ is nondecreasing and
$\lim_{r\to 0^+}\tilde \alpha(r)=0$ there exists $\hat \alpha \in \Ki$ such
that $\hat \alpha(r)\ge \tilde \alpha(r)$ for all $r\ge 0$. Let
$x \in \T(t_0,u,\sigma)$ with $t_0\ge 0$, $u\in \U_m$ and
$\sigma\in \S$. Let $t\ge t_0$, and let $u_t$ be the input $u_t(\tau)=u(\tau)$
for all $\tau \in [t_0,t]$ and $u_t(\tau)=0$ otherwise. From
well-known results on differential equations, there exists
$x^*\in \T(t_0,u_t,\sigma)$ such that $x^*(\tau)=x(\tau)$ for all
$\tau \in [t_0,t]$. By using the definition of $\tilde \alpha$ and the
facts that $\|u_t\|=\int_{t_0}^t\chi(|u(s)|)\:ds$ and
$\hat \alpha(r)\ge \tilde \alpha(r)$, we then have
\begin{align*}
  |x(t)|=|x^*(t)| &\le \hat \alpha(|x(t_0)|)+ \hat \alpha\left (\|u_t\|\right )\\
  &= \hat \alpha(|x(t_0)|)+\hat \alpha \left (\int_{t_0}^t\chi(|u(s)|)\:ds \right ).
\end{align*}
In consequence, the lemma follows by taking
$\tilde \alpha_1=\tilde \alpha_2 =\hat \alpha$.\hfill\QED

\subsection{Proof of Lemma~\ref{lem:ope}}
\label{sec:pf-lem-ope}
Let $\alpha$ be continuous and positive definite. We will prove that $(\hat{h},f)$, with $\hat{h}=\alpha(|h|)$, is zero-input output-PE w.r.t. $\S$ by contradiction. Suppose that $(\hat{h},f)$ is not zero-input output-PE w.r.t. $\S$. Let $\hat{h}_0(t,\xi,i)\equiv \hat{h}(t,\xi,0,i)$. Then there exist $\varepsilon_0>0$ and a sequence $\{(t_k,x_k,\sigma_k)\}$ such that $t_k\nearrow \infty$ and, for all $k$, $\sigma_k\in \S$  and
$x_k\in \T(t_k,\mathbf{0},\sigma_k)$, $\varepsilon_0\le |x_k(t)|\le 1/\varepsilon_0$ for all $t\in [t_k,t_k+k]$, and 
 $$ \int_{t_k}^{t_{k}+k} |\hat{h}_0(\tau,x_k(\tau),\sigma_k(\tau)|^2\:d\tau<1/k.$$
 Let $\tilde y_k(s)=|\hat{h}_0(t_k+s,x_k(t_k+s),\sigma_k(t_k+s)|^2$ for $s\in [0,k]$ and $\tilde y_k(s)=0$ if $s>k$. Since $\{\tilde y_k\}$ converges to $0$ in $L^1(\R_{\ge 0})$, then there exists a subsequence $\{\tilde y_{k_l}\}$ such that $\lim_{l\to \infty}\tilde y_{k_l}(s)=0$ for almost all $s \in \R_{\ge 0}$. The fact that $h_0$ is essentially bounded on $\R_{\ge 0}\times K \times \Gamma$, with $K=\{\xi\in \R^n:\varepsilon_0\le |\xi|\le 1/\varepsilon_0\}$,  implies the existence of a constant $M\ge 0$ so that $|h_0(t_{k_l}+s,x_{k_l}(t_{k_l}+s),\sigma_{k_l}(t_{k_l}+s)|\le M$ for almost all $s\in [0,k_l]$. From the latter, the continuity and positive definiteness of $\alpha$ and the fact that $\tilde y_{k_l}\to 0$ a.e., we have that for every $T>0$, $\lim_{l\to \infty} h_0(t_{k_l}+s,x_{k_l}(t_{k_l}+s),\sigma_{k_l}(t_{k_l}+s))=0$ for almost all $s\in [0,T]$. By applying Lebesgue's Convergence Theorem it follows that for all $T>0$,
 $$ \lim_{l\to \infty} \int_0^T| h_0(t_{k_l}+s,x_{k_l}(t_{k_l}+s),\sigma_{k_l}(t_{k_l}+s)|^2\: ds=0,$$
 or, equivalently, that
  $$ \lim_{l\to \infty} \int_{t_{k_l}}^{t_{k_l}+T}| h_0(\tau,x_{k_l}(\tau),\sigma_{k_l}(\tau)|^2\: d\tau=0.$$
We have arrived to a contradiction because from the zero-input output-PE w.r.t. $\S$ of the pair $(h,f)$ there exist $T(\varepsilon_0)>0$ and $r(\varepsilon_0)>0$ such that for all $l$
$$ \int_{t_{k_l}}^{t_{k_l}+T(\varepsilon_0)}| h_0(\tau,x_{k_l}(\tau),\sigma_{k_l}(\tau)|^2\: ds\ge r(\varepsilon_0).$$
\hfill \QED
\bibliographystyle{IEEEtran}
\bibliography{/home/hhaimo/latex/strings.bib,/home/hhaimo/latex/complete_v2.bib,/home/hhaimo/latex/Publications/hernan_v2.bib}
\end{document}